\documentclass[sigconf]{acmart}
\setcopyright{none}
\renewcommand\footnotetextcopyrightpermission[1]{} 

\usepackage{amsmath,amsfonts}
\usepackage[utf8]{inputenc}
\usepackage{graphicx}
\usepackage{algorithm}
\usepackage{algpseudocode}
\usepackage{caption}
\usepackage{subcaption}
\usepackage{booktabs}
\usepackage{mathtools}
\usepackage{threeparttable}
\usepackage{caption}
\usepackage{subcaption}
\usepackage{todonotes}
\usepackage{mathtools}
\usepackage{etoolbox}  
\usepackage{array}
\usepackage{dblfloatfix}

\newcolumntype{L}[1]{>{\raggedright\let\newline\\\arraybackslash\hspace{0pt}}m{#1}}
\newcolumntype{C}[1]{>{\centering\let\newline\\\arraybackslash\hspace{0pt}}m{#1}}
\newcolumntype{R}[1]{>{\raggedleft\let\newline\\\arraybackslash\hspace{0pt}}m{#1}}

\AtBeginDocument{%
  \providecommand\BibTeX{{%
    \normalfont B\kern-0.5em{\scshape i\kern-0.25em b}\kern-0.8em\TeX}}}

\setcopyright{acmcopyright}
\copyrightyear{2018}
\acmYear{2018}
\acmDOI{XXXXXXX.XXXXXXX}

\acmConference[SEAMS 2022]{The 17th Symposium on Software Engineering for Adaptive and Self-Managing Systems}{May 21–29, 2022}{Pittsburgh, PA, USA}
\acmPrice{15.00}
\acmISBN{978-1-4503-XXXX-X/18/06}

\newcommand{\squishlist}{
   \begin{list}{$\bullet$}
    { \setlength{\itemsep}{2pt}    \setlength{\parsep}{0pt}
      \setlength{\topsep}{5pt}     \setlength{\partopsep}{0pt}
      \setlength{\leftmargin}{1.35em} \setlength{\labelwidth}{1em}
      \setlength{\labelsep}{0.5em} } }

\newcommand{\squishend}{
    \end{list}  }

\begin{document}

\title{PRESTO: Predicting System-level Disruptions through Parametric Model Checking}

\author{Xinwei Fang}
\affiliation{%
  \institution{Department of Computer Science}
  \institution{University of York \city{York} \country{UK}}
}
\email{xinwei.fang@york.ac.uk}

\author{Radu Calinescu}
\affiliation{%
  \institution{Department of Computer Science}
  \institution{University of York \city{York} \country{UK}}
}
\email{radu.calinescu@york.ac.uk}

\author{Colin Paterson}
\affiliation{%
  \institution{Department of Computer Science}
  \institution{University of York \city{York} \country{UK}}
}
\email{colin.paterson@york.ac.uk}

\author{Julie Wilson}
\affiliation{%
  \institution{Department of Mathematics}
  \institution{University of York \city{York} \country{UK}}
}
\email{julie.wilson@york.ac.uk}


\begin{abstract}
Self-adaptive systems are expected to mitigate disruptions by continually adjusting their configuration and behaviour. This mitigation is often reactive. Typically, environmental or internal changes trigger a system response only after a violation of the system requirements. Despite a broad agreement that prevention is better than cure in self-adaptation, proactive adaptation methods are underrepresented within the repertoire of solutions available to the developers of self-adaptive systems. To address this gap, we present a work-in-progress approach for the \underline{pre}diction of \underline{s}ystem-level disrup\underline{t}i\underline{o}ns (PRESTO) through parametric model checking. Intended for use in the analysis step of the MAPE-K (Monitor-Analyse-Plan-Execute over a shared Knowledge) feedback control loop of self-adaptive systems, PRESTO comprises two stages. First, time-series analysis is applied to monitoring data in order to identify trends in the values of individual system and/or environment parameters. Next, future non-functional requirement violations are predicted by using parametric model checking, in order to establish the potential impact of these trends on the reliability and performance of the system. We illustrate the application of PRESTO in a case study from the autonomous farming domain.
\end{abstract}





\maketitle

\section{Introduction}
Self-adaptive systems are expected to mitigate disruptions in complex and uncertain environments by continually analysing, evaluating and adjusting their configurations. To achieve this, their behaviours and operating environments need to be modelled and analysed, often using stochastic models such as queueing networks~\cite{4620121}, Markov models~\cite{gerasimou2018synthesis,paterson2018observation} and stochastic Petri nets~\cite{balsamo2012methodological}. These models are continually updated to reflect changes observed through monitoring, and used to re-verify compliance with the requirements when needed. In this work, we focus on the use of discrete-time Markov chains (DTMCs) for such purpose and consider disruptions as violations of system-level requirements.

Proactive adaptations that make  decisions before disruptions occur have a number of advantages over  traditional reactive adaptations, especially when associated with cost~\cite{poladian2007leveraging} or latency~\cite{moreno2015proactive}. However, many proactive adaptations are triggered by the changes of system parameters rather than the anticipated violation of system-level properties~\cite{amin2012automated,metzger2012predictive}. Due to the complexity and non-linearity of system processes, the impact at system-level caused by these low-level changes may not be obvious. For example, large changes in system parameters not necessarily lead to the violation of requirements but small ones may ~\cite{fisler2005verification}. Therefore, adaptation decisions triggered by changes in system parameters may result in over or under adaptations.     

Fortunately, recent advances in the analysis of \emph{parametric} discrete-time Markov chains allow the efficient verification of system-level requirements using runtime observations of system parameters~\cite{Jansen2014,RaduePMC,XinweiICSE21}. Additionally, a growing collection of methods for online learning the parameters of these models ensures that underlying changes or trends in these parameters can be successfully captured~\cite{calinescu2014adaptive,filieri2015lightweight,zhao2020interval}. Finally, advances in data analysis allow the confident prediction of observed parameters~\cite{metzger2020triggering,metzger2019proactive,amin2012automated}. Combining these methods can enable self-adaptive systems to \emph{predict and prevent} disruptions before they happen.

Our paper introduces a work-in-progress approach for the \underline{pre}dic\-tion of \underline{s}ystem-level disrup\underline{t}i\underline{o}ns (PRESTO) through parametric model checking. PRESTO is intended for use in the analysis step of the MAPE-K (Monitor-Analyse-Plan-Execute over a shared Knowledge) feedback control loop~\cite{brun2009engineering,arcaini2015modeling} of self-adaptive systems. PRESTO allows adaptation decisions to be made based on predicted system-level requirement violations due to  degradation in system and/or environmental parameters---a common phenomenon in real-world applications~\cite{sankavaram2009model,russell2021stochastic} that is underexplored by current research into self-adaptive systems. PRESTO can work in conjunction with methods that mitigate other types of disruptions in self-adaptive systems, e.g., sudden requirement violations due to step changes in system and/or environment parameter values ~\cite{zhao2020interval,filieri2012formal}.

The rest of paper is structured as follows. In Section~\ref{preliminaries}, we provide a formal definition for Markov models and how to define and evaluate system-level properties. Section~\ref{Sec:runningexample} gives a running example that will be used throughout this paper. We present PRESTO and its preliminary evaluation in Section~\ref{sec:method} and~\ref{sec:evaluation}, respectively, followed by related work on proactive adaptation in Section~\ref{sec:relatedwork}. Lastly, we conclude the paper with a brief summary in Section~\ref{sec:conclusion}. 


\section{Preliminaries}
\label{preliminaries}

\textbf{Discrete-time Markov chains (DTMCs)} are finite state-transition systems comprising \textit{states} associated with relevant configurations of a system under analysis, and \emph{probabilistic state transitions} that model the stochastic behaviour of that system.



\begin{definition}
A discrete-time Markov chain over a set of atomic propositions $\mathit{AP}$ is a tuple
\begin{equation}
    \label{eq:dtmc}
    D=(S, s_{init}, M, L), 
\end{equation}
where 
$S\neq\emptyset$ is a finite set of states; $s_{init} \in S$ is the initial state; $M: S \times S \rightarrow [0,1]$ is a transition probability matrix such that, for any pair of states $s, s' \in S, M(s, s')$ provides a value indicating the probability of transitioning from $s$ to $s'$, and $\sum_{s' \in S} M(s,s')=1$ for any $s \in S$; $L: S \rightarrow 2^\mathit{AP}$ is a labelling function that maps every state $s \in S$ to the atomic propositions from $\mathit{AP}$ that hold in that state. 
\end{definition}

\smallskip\noindent
 A \textit{reward} can be assigned to states to extend the range of non-functional properties that can be analysed. 

\begin{definition}
A reward structure over a discrete-time Markov chain~\eqref{eq:dtmc} is a function
\begin{equation}
    \label{eq:reward}
    \mathit{rwd}:S \rightarrow \mathbb{R}_{\geq 0}
\end{equation}
that associates non-negative values with the DTMC states.
\end{definition}

\smallskip\noindent
Parametric discrete-time Markov chains (pDTMCs) can be utilised for the analysis of reward-augmented DTMCs with unknown probabilities and/or rewards. 

\begin{definition}
\label{def:pmc}
A parametric discrete-time Markov chain (pDTMC) is a discrete-time Markov chain~\eqref{eq:dtmc}, with or without a set of reward functions~\eqref{eq:reward}, that includes transition probabilities $P(s,s')$ and/or rewards specified as rational functions over a set of real-valued parameters. 
\end{definition}

\noindent
\textbf{Probabilistic computation tree logic (PCTL)}~\cite{bianco_alfaro_1995,hansson1994logic,andova2003discrete} is a probabilistic variant of temporal logic used to specify the properties of DTMCs.

\begin{definition}
A PCTL \emph{state formula} $\Phi$, \emph{path formula} $\Psi$, and \emph{reward state formula} $\Phi_\mathsf{R}$ over an atomic proposition set $\mathit{AP}$ are defined by:

    \begin{equation}
    \label{FormState}
    \Phi::=  true \,\vert\, a \,\vert\, \neg \Phi \,\vert\, \Phi \wedge \Phi \,\vert\, \mathcal{P}_{\!=?} [\Psi] \\[0.5mm]
    \end{equation}
    \begin{equation}
    \label{FormPath}
    \Psi::= \mathrm{X} \Phi \;\vert\; \Phi\; \mathrm{U}^{\leq k}\, \Phi \vert\; \Phi\; \mathrm{U}\; \Phi \; \\[0.5mm]
    \end{equation}
    \begin{equation}
    \label{FormReward}
    \Phi_\mathsf{R}^::= 
    \mathcal{R}_{\!=?}^\mathit{rwd}[\mathrm{I}^{=k}] \;\vert\; \mathcal{R}_{\!=?}^\mathit{rwd}[\mathrm{C}^{\leq k}] \;\vert\;
    \mathcal{R}_{\!=?}^\mathit{rwd}[\mathrm{F}\; \Phi] \;\vert\; 
    \mathcal{R}_{\!=?}^\mathit{rwd}[\mathrm{S}]
    \end{equation}

\noindent
where $a \in AP$ is an atomic proposition, $k \in \mathbb{N}_{>0}$ is a timestep bound, and $\mathit{rwd}$ is a reward structure.
\end{definition}

The PCTL semantics is defined using a satisfaction relation $\models$ over the states $s\in S$ and paths $\pi\in \mathit{Paths}^D(s)$ of a DTMC. Thus, $s\models \Phi$ means ``$\Phi$ holds in state $s$'', $\pi\models \Psi$ means ``$\Psi$ holds for path $\pi$'', and we have: $s\models true$ for all states $s\in S$; $s \models a$ iff $a\in L(s)$;$s \models \neg \Phi$ iff $\neg (s\models \Phi)$; $s\models \Phi_1 \wedge \Phi_2$ iff $s\models \Phi_1$ and $s\models \Phi_2$. The quantitative state formula $\mathcal{P}_{\!=?} [\Psi]$ specifies the probability that paths from $\mathit{Paths}^D\!(s)$ satisfy the path property $\Psi$. \emph{Reachability properties}  $\mathcal{P}_{\!=?} [\mathsf{true}\, \mathrm{U}\, \Phi]$ are equivalently written as $\mathcal{P}_{\!=?} [\mathsf{F}\, \Phi]$ or $\mathcal{P}_{\!=?} [\mathsf{F}\, R]$, where $R\!\subseteq\! S$ is the set of states in which $\Phi$ holds. The \emph{next formula} $X \Phi$ holds if $\Phi$ is satisfied in the next state of the analysed path $\pi$. The \emph{time-bounded until formula} $\Phi_1\, \mathrm{U}^{\leq k}\, \Phi_2$ holds for a path $\pi$ iff $\pi(i)\models \Phi_2$ for some $i\leq k$ and $\pi(j)\models \Phi_1$ for all $j=1,2,\ldots,i-1$. The \emph{unbounded until formula} $\Phi_1\,\mathrm{U}\, \Phi_2$ removes the bound $k$ from the time-bounded until formula. Finally, the reward state formulae specify the expected values for: the \emph{instantaneous reward} at timestep $k$ ($\mathcal{R}_{\!=?}^\mathit{rwd}[\mathrm{I}^{=k}]$); the \emph{cumulative reward} up to timestep $k$ ($\mathcal{R}_{\!=?}^\mathit{rwd}[\mathrm{C}^{\leq k}]$); the \emph{reachability reward} cumulated until reaching a state that satisfies a property $\Phi$ ($\mathcal{R}_{\!=?}^\mathit{rwd}[\mathrm{F}\; \Phi]$); The \emph{steady-state reward} in the long run ($\mathcal{R}_{\!=?}^\mathit{rwd}[\mathrm{S}]$). For detailed descriptions of the PCTL semantics, see~\cite{bianco_alfaro_1995,hansson1994logic,andova2003discrete}.

\smallskip\noindent
\textbf{Parametric model checking (PMC)}~\cite{daws2004symbolic} is a mathematically based technique for the verification of PCTL-encoded pDTMC properties. Supported by probabilistic model checkers such as PRISM~\cite{prism} and Storm~\cite{storm}, the technique yields rational functions (i.e., a quotients between two polynomials with respect to the parameters) for the analysed PCTL property. These functions, which we term \emph{PMC expressions} in the remainder of the paper, can be efficiently evaluated by a self-adaptive system at runtime, when the value of the pDTMC parameters are observed through the MAPE-K monitoring. 

\begin{figure}
    \centering
    \includegraphics[width=0.42 \textwidth]{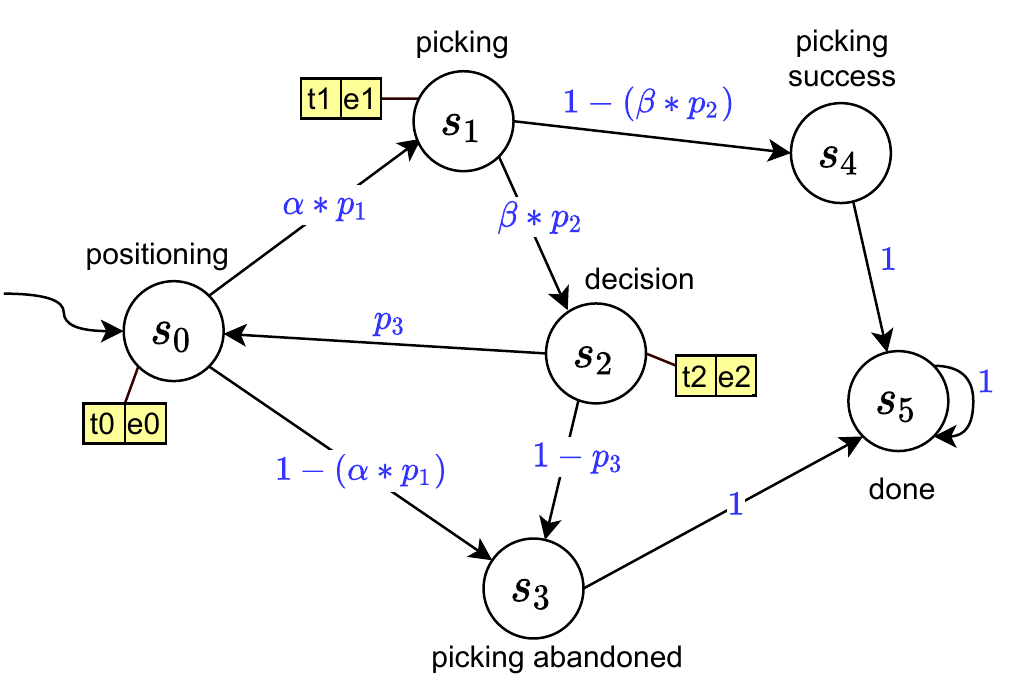}
    \vspace*{-0.2cm}
    \caption{pDTMC model of autonomous fruit-picking process}
    \label{fig:pDTMCexample}

    \vspace*{-0.2cm}
\end{figure}

\begin{table*}[]
    \centering
    \sffamily
    \begin{small}
     \caption{System-level requirements, related properties, and the result of parametric model checking for the properties}
     \label{Propsummary}
     
    \vspace*{-2mm}
    \begin{tabular}{C{0.3cm} p{6cm} p{4cm} p{6cm}}
    \toprule
     {}& \textbf{Informal description} & \textbf{Property to check (PCTL)} & \textbf{PMC expressions$\dagger$} \\ \midrule
    R1 &  The robot shall complete the fruit picking successfully with probability of at least 0.8 & $\mathcal{P}_{\!=?}[\mathrm{F}$ ``picking success''$] \geq 0.8$ & $(\alpha*p1*\beta*p2+(-1)*\alpha*p1)/(\alpha*p1*\beta*p2*p3+(-1))$ \\ 
    \midrule
    R2 &  The expected time to complete the picking process shall not exceed 30 seconds. &$\mathcal{R}_{\!=?}^\mathit{``time"}[\mathrm{F}]$``done" $] \leq 30s$ & $(-1 * (\alpha*p1*\beta*p2*t3+t1+\alpha*p1*t2))/(\alpha*p1*\beta*p2*p3+(-1))$ \\ 
    \midrule
    R3 &  The expected energy consumption to complete the picking process shall not exceed 10 joules. &$\mathcal{R}_{\!=?}^\mathit{``energy"}[\mathrm{F}]$``done" $] \leq 10J$ & $(-1 * (\alpha*p1*\beta*p2*e3+e1+\alpha*p1*e2))/(\alpha*p1*\beta*p2*p3+(-1))$ \\ 
    \bottomrule
    \end{tabular}
    \end{small}
    \raggedright
    $^\dagger$\footnotesize PMC expressions were returned from the model checker Storm~\cite{storm} by providing the pDTMC and the property specified in PRISM language~\cite{prism} and PCTL respectively.  
    
    \vspace*{-2mm}
\end{table*}

\begin{figure*}[b]
	\centering
    \includegraphics[width=0.86\textwidth]{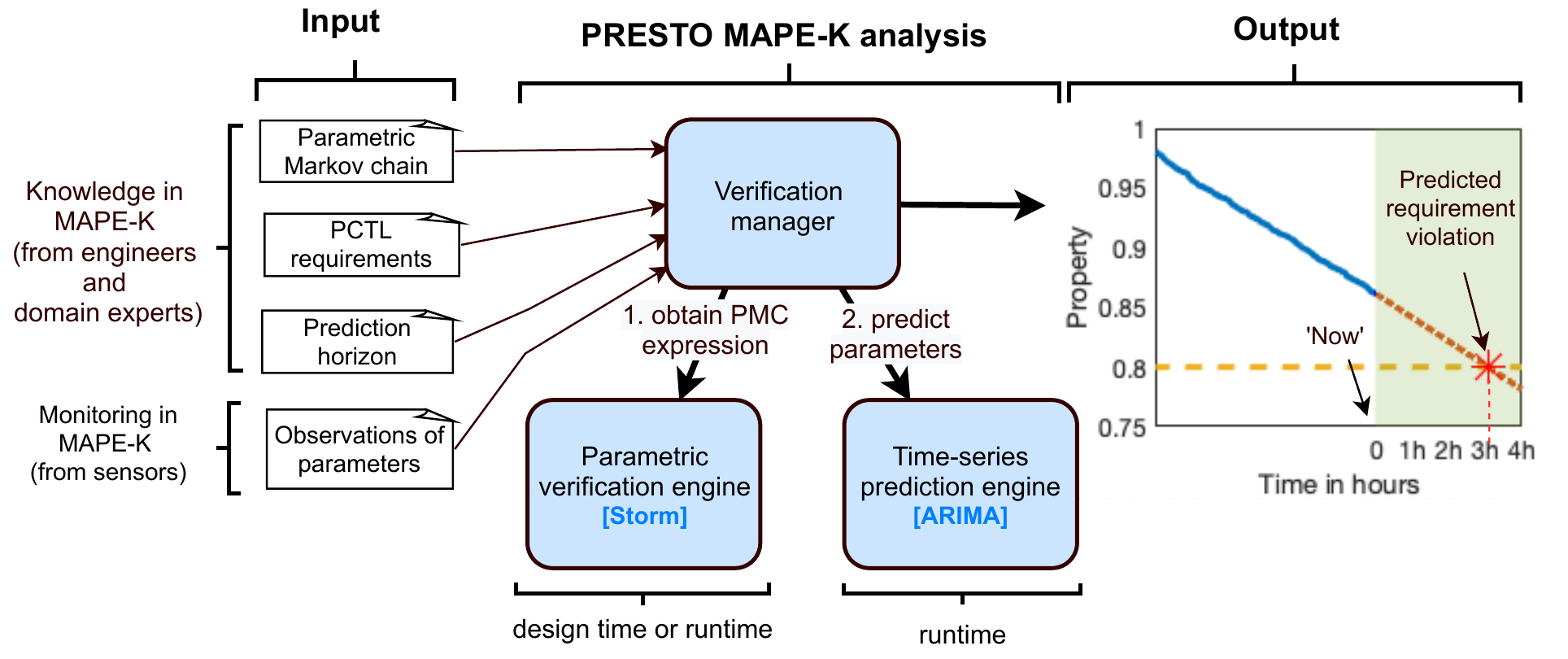}
    
    \vspace*{-2mm}
    \caption{PRESTO analysis process and architecture}
    \label{Overview}
\end{figure*}

\section{Motivating example}
\label{Sec:runningexample}
We motivate PRESTO and demonstrate its effectiveness for a fruit-picking robot
from the autonomous farming domain.
Our example, inspired by recent development in this domain~\cite{wagner2021efficient,xiong2021improved}, assumes that the robot needs to perform three operations autonomously:
(1)~position itself in the right location for the fruit picking; (2)~use its arm to pick up the fruit; and (3)~when operation~2 is unsuccessful, decide whether to retry the fruit picking from operation~1 or to give up. 
Fig.~\ref{fig:pDTMCexample} shows the pDTMC model of this process.
The robot begins with positioning itself next to a piece of fruit (state $s_0$). The positioning operation may succeed, in which case the robot moves to state $s_1$, where it attempts picking, or may fail, in which case the picking will be abandoned ($s_3$) and the process finishes ($s_5$). The process also finishes when the picking is successful ($s_4$). If the picking is unsuccessful, the robot enters the decision state ($s_2$) from where it needs to decide whether to re-position itself and retry the entire process, or to abandon picking this fruit ($s_3$) and end the process ($s_5$). 

As shown in Fig.~\ref{fig:pDTMCexample}, the outgoing transition probabilities from states $s_0$, $s_1$ and $s_2$ are defined in terms of several system parameters. Out of these parameters, we assume that $p_1$, $p_2$ and $p_3$ have predefined, domain-specific values. For instance, $p_1$ is the probability of successful positioning when the robot is in perfect working condition. In contrast, $\alpha,\beta\in(0,1]$ represent coefficients that reflect the robot degradation from its perfect working condition, e.g. due to wear and tear. We assume that $\alpha$ and $\beta$ are unknown, and their values need to be obtained through monitoring~\cite{li2019joint} or via self-testing~\cite{paschalis2004effective}.

Additionally, the pDTMC is annotated with two reward functions (depicted in rectangular boxes linked to states $s_0$, $s_1$ and $s_2$ from Fig.~\ref{fig:pDTMCexample}). First, a ``time'' reward function associates mean operation execution times $t_0$, $t_1$ and $t_2$ with the three operations performed by the robot. Similarly, an ``energy'' reward function associates mean energy consumptions $e_0$, $e_1$ and $e_2$ with the same operations. We assume that the values of $t_i$ and $e_i$, $0\leq i\leq 2$, are unknown and need to be obtained through monitoring.


Finally, we assume that the autonomous robot has to satisfy the three system-level requirements shown in 
Table~\ref{Propsummary}.

\begin{figure*}[b]
     \centering
     \begin{subfigure}[b]{0.32\textwidth}
         \centering
         \includegraphics[width=\textwidth]{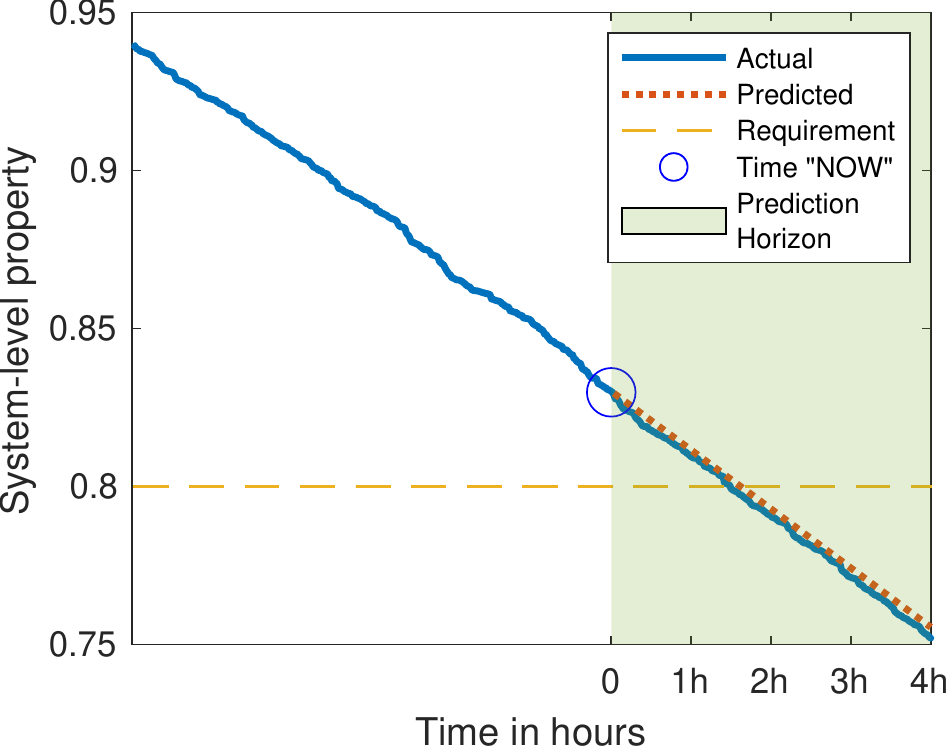}
         \caption{$R1$}
         \label{fig:RQ1R1}
     \end{subfigure}
     \hfill
     \begin{subfigure}[b]{0.32\textwidth}
         \centering
         \includegraphics[width=\textwidth]{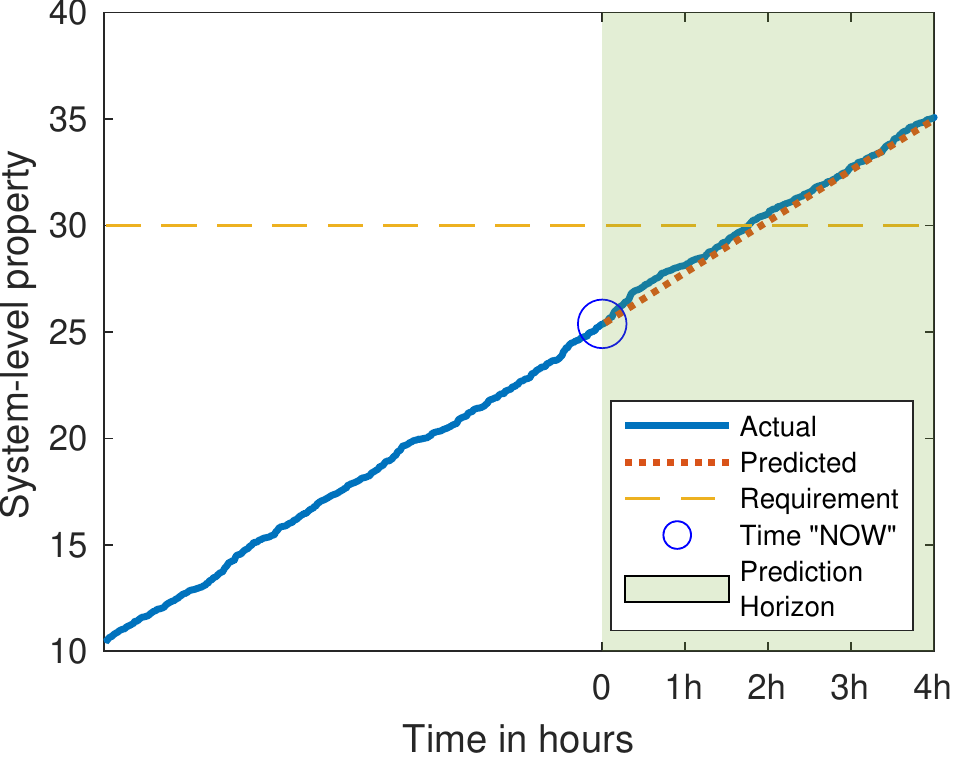}
         \caption{$R2$}
         \label{fig:RQ1R2}
     \end{subfigure}
     \hfill
     \begin{subfigure}[b]{0.32\textwidth}
         \centering
         \includegraphics[width=\textwidth]{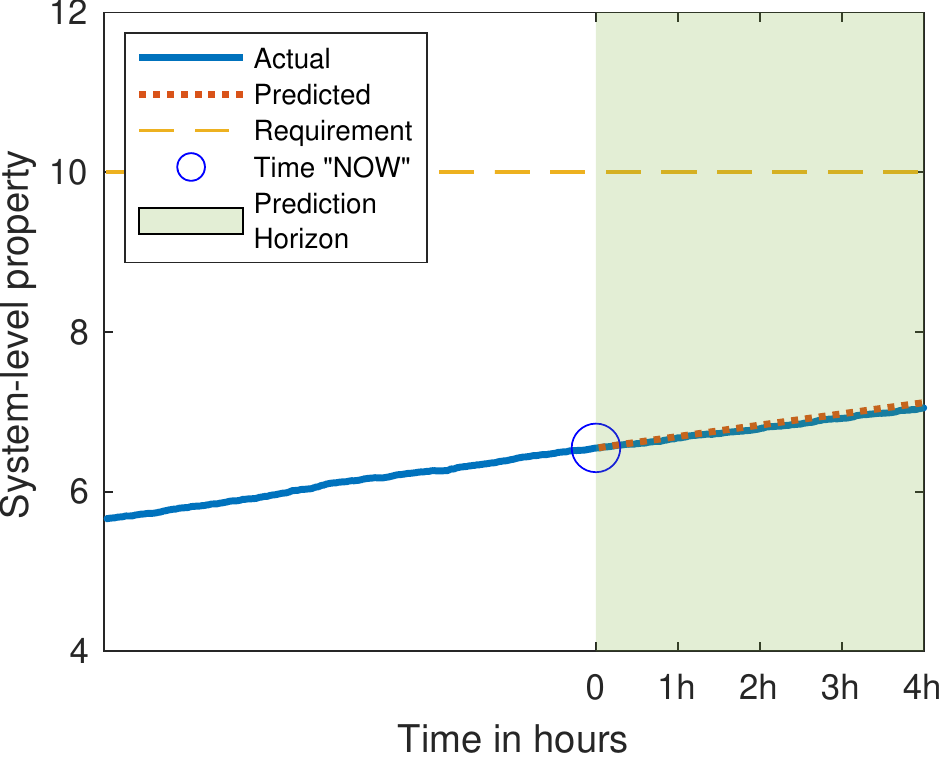}
         \caption{$R3$}
         \label{fig:RQ1R3}
     \end{subfigure}
        \caption{Prediction of system-level properties and potential disruptions for the properties listed in Table~\ref{Propsummary}, the actual disruptions happen at 89-min (R1) and 105-min(R2) in contrast to the predicted disruptions are at 97-min(R1) and 115-min(R2), respectively}
        \label{fig:eva1}
\end{figure*}

\section{PRESTO approach}
\label{sec:method}

The PRESTO analysis process, depicted in Fig.~\ref{Overview}, takes as input a pDTMC model of the self-adaptive system's behaviour, a set of PCTL-encoded system requirements, a \emph{prediction horizon} $h>0$, and runtime observations of the unknown system and/or environment parameters that appear in the pDTMC. The first three inputs are provided, at design time, by system engineers and domain experts, whereas the runtime observations are obtained from the monitoring component of the MAPE-K control loop. 

Given these inputs, PRESTO predicts 
violation of the system-level requirements \emph{within the prediction horizon} $h$. To that end, our approach first uses a parametric verification engine to obtain PMC expressions for the PCTL properties associated with the system requirements. Typically, this is a one-off step executed at design time; however, PRESTO also allows the run-time execution of this step, if needed because the model structure and/or the system requirements change in operation. The PMC expressions are then used at runtime, to continually assess the impact of predicted parameter changes on the system-level requirements. In this step, PRESTO uses recent observations of the parameter values to predict the parameter future values within the prediction horizon. These predicted parameter values are then ``propagated'' through the PMC expressions (i.e., they are used to evaluate the PMC expressions) in order to predict the future values of the nonfunctional properties from the system requirements, and thus to predict future requirement violations.

\section{Evaluation}
\label{sec:evaluation}
A prototype of PRESTO was implemented to aid evaluation and answer the following research questions: 
\begin{itemize}
    \item[RQ1] Can PRESTO predict the violation of system-level disruptions and how does it perform in terms of prediction errors?
    \item[RQ2] How does the noise level in model parameters affect PRESTO's prediction performance?
    \item[RQ3] How can we utilise PRESTO for proactive adaptation?
\end{itemize}

\smallskip\noindent
The prototype and a simple simulator of the autonomous robot from our motivating example were implemented in Java, using a modular design as shown in Fig.~\ref{Overview}. The modular design enables: 1) technologies used in this version of PRESTO to be replaced with alternatives if deemed more suitable for a given application and 2) additional technologies can be integrated to extend the functionality or to improve efficiency. This early version of PRESTO uses the following technologies:


\squishlist
    \item The parametric verification engine uses Storm ~\cite{storm}, one of the leading parametric model checkers, in the background. As an alternative, fPMC~\cite{XinweiICSE21} is worth exploring, especially when the system under analysis is complex and with many parameters.   
    \item The time-series prediction engine uses ARIMA (Autoregressive Integrated Moving Average), one of the most widely used methods for time-series prediction, also invoked in the background. Numerous alternative prediction methods could be substituted for this module depending on requirements. 
\squishend



\begin{figure*}
     \centering
     \begin{subfigure}[b]{0.32\textwidth}
         \centering
         \includegraphics[width=\textwidth]{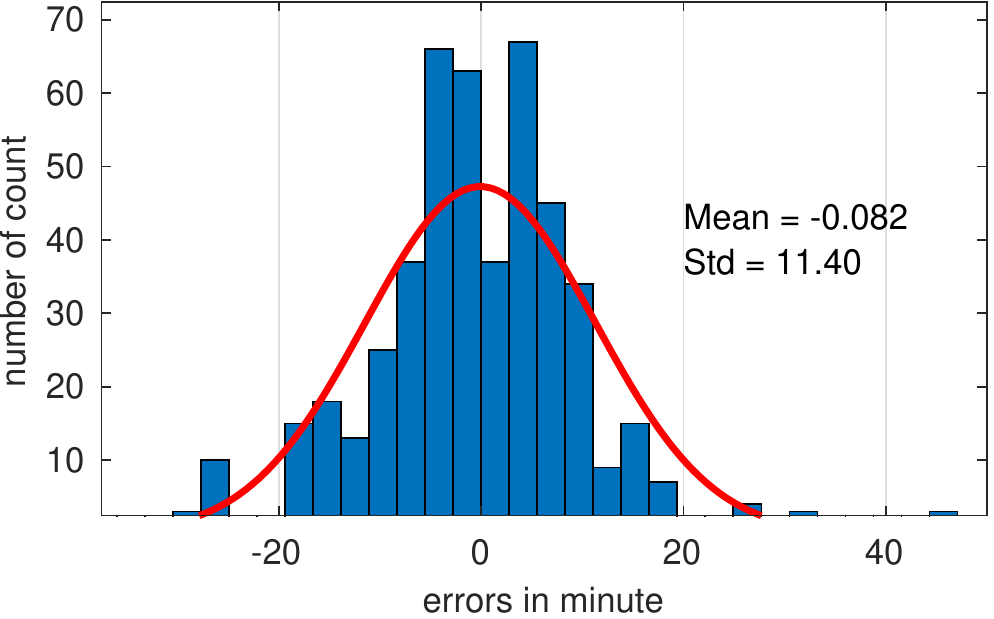}
         \caption{$R1$}
         \label{fig:RQ2R1}
     \end{subfigure}
     \hfill
     \begin{subfigure}[b]{0.31\textwidth}
         \centering
         \includegraphics[width=\textwidth]{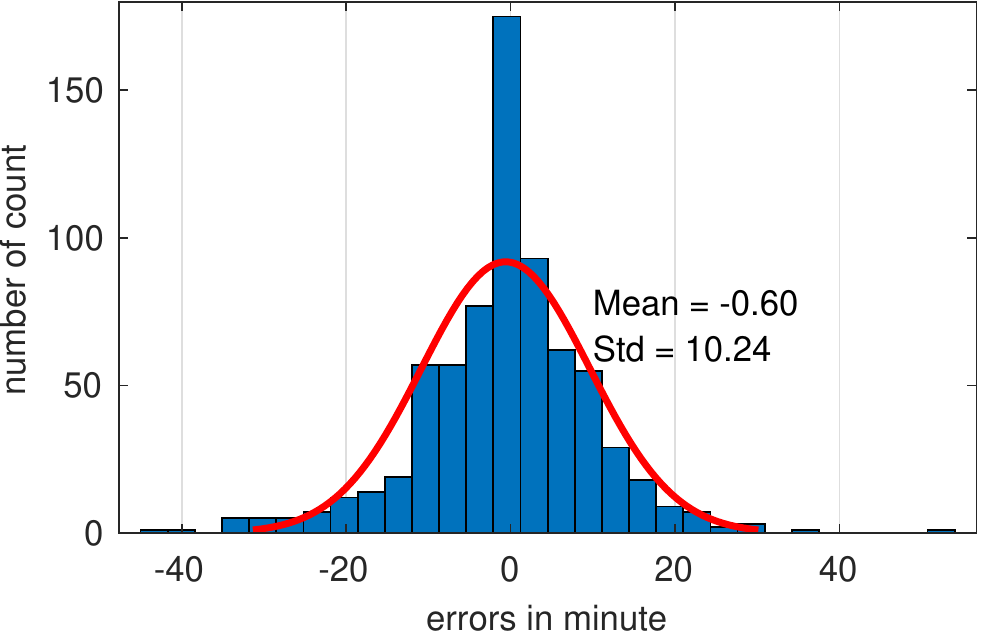}
         \caption{$R2$}
         \label{fig:RQ2R2}
     \end{subfigure}
     \hfill
     \begin{subfigure}[b]{0.32\textwidth}
         \centering
         \includegraphics[width=\textwidth]{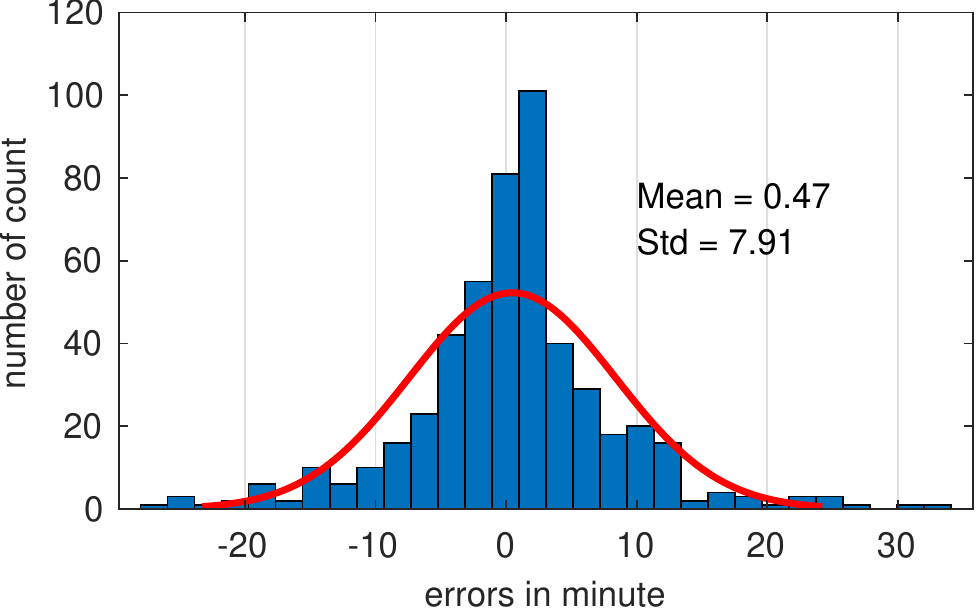}
         \caption{$R3$}
         \label{fig:RQ2R3}
     \end{subfigure}
        \caption{Errors of predicting violation of requirements collected from experiments that have violation of requirements within the prediction horizon}
        \label{fig:eva2}
\end{figure*}

\begin{figure*}
     \centering
     \begin{subfigure}[b]{0.32\textwidth}
         \centering
         \includegraphics[width=\textwidth]{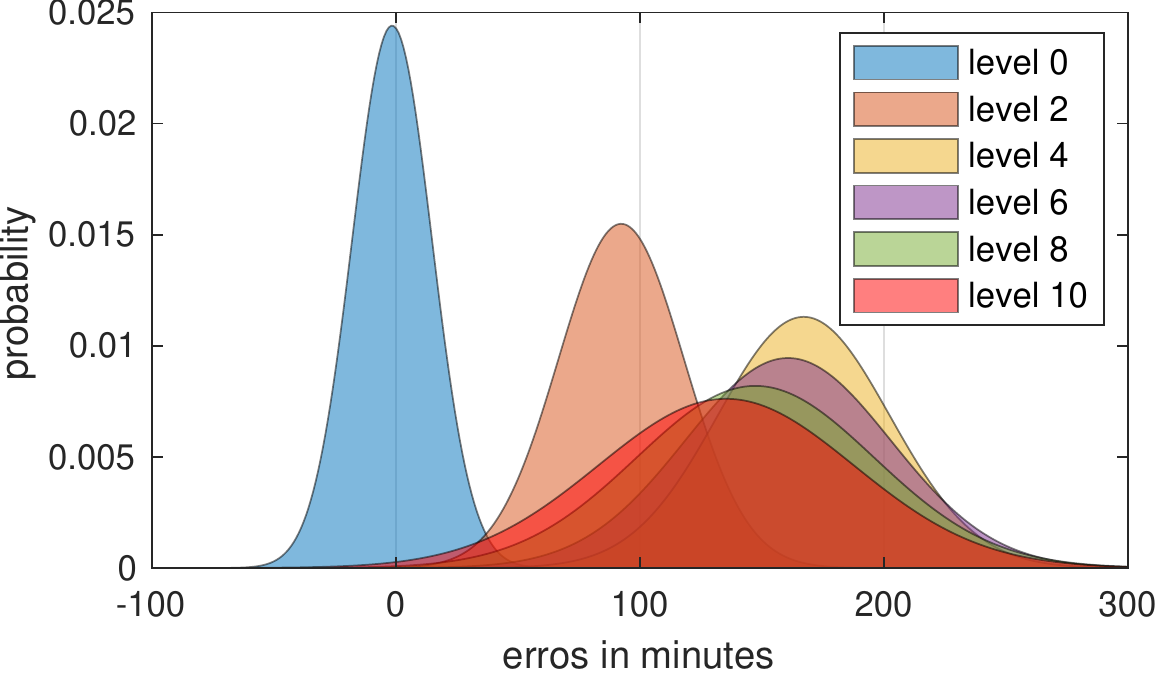}
         \caption{$R1$}
         \label{fig:RQ3R1}
     \end{subfigure}
     \hfill
     \begin{subfigure}[b]{0.32\textwidth}
         \centering
         \includegraphics[width=\textwidth]{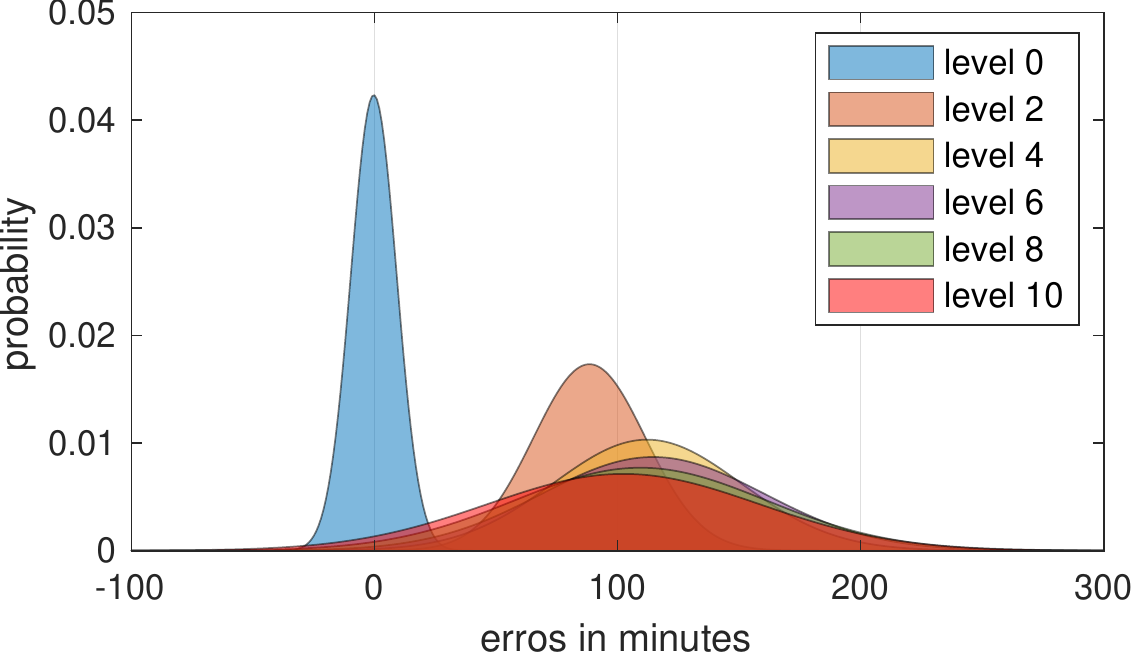}
         \caption{$R2$}
         \label{fig:RQ3R2}
     \end{subfigure}
     \hfill
     \begin{subfigure}[b]{0.32\textwidth}
         \centering
         \includegraphics[width=\textwidth]{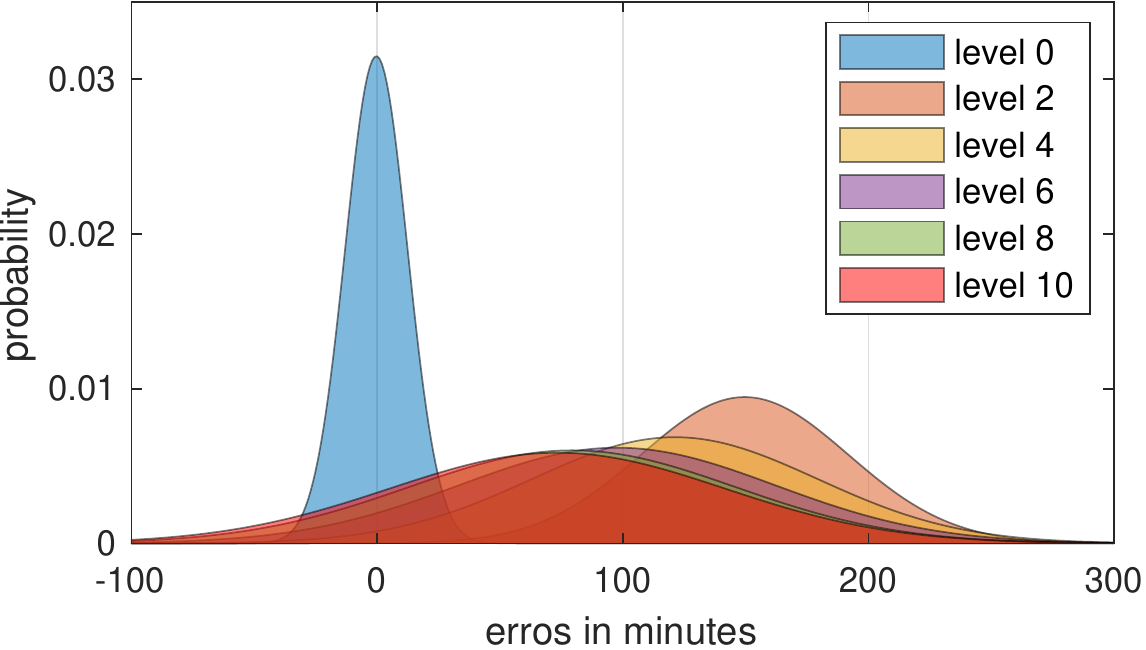}
         \caption{$R3$}
         \label{fig:RQ3R3}
     \end{subfigure}
        \caption{Changes of error distributions as the results of noise level in parameters}
        \label{fig:eva3}
\end{figure*}

\begin{figure*}
     \centering
     \begin{subfigure}[b]{0.32\textwidth}
         \centering
         \includegraphics[width=0.8\textwidth]{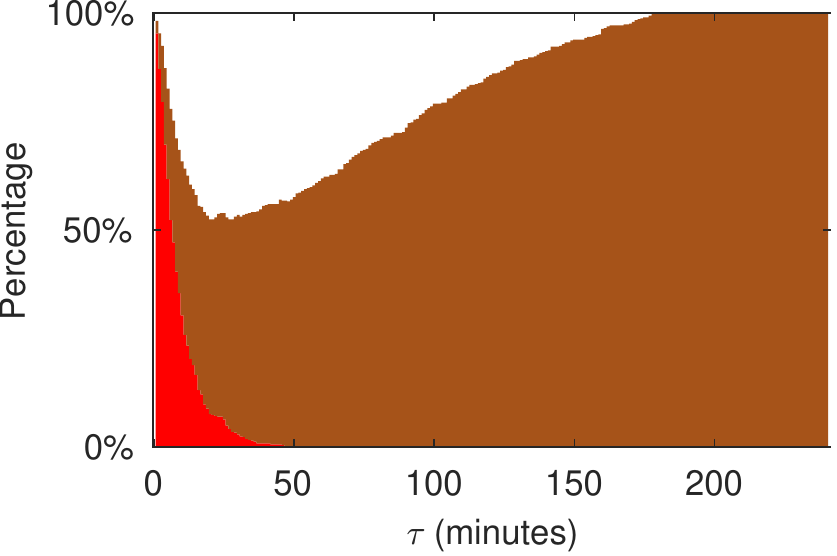}
         \caption{$R1$}
         \label{fig:TauR1}
     \end{subfigure}
     \hfill
     \begin{subfigure}[b]{0.32\textwidth}
         \centering
         \includegraphics[width=0.8\textwidth]{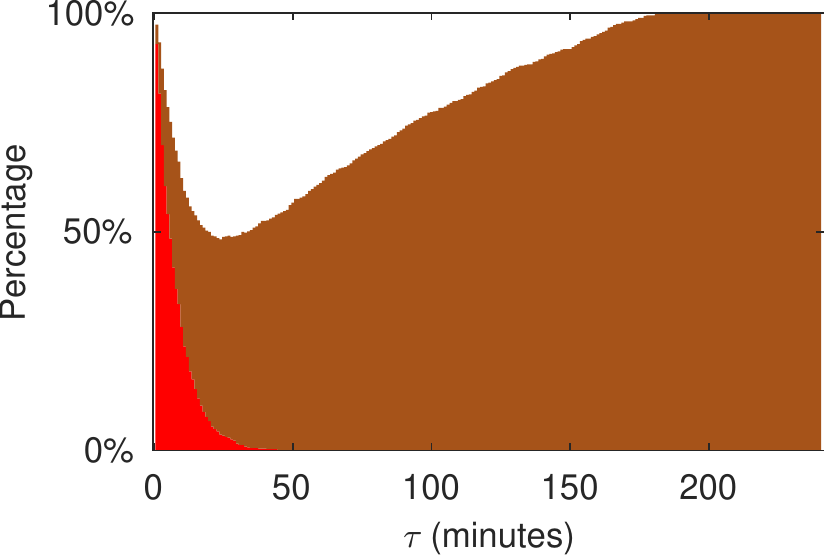}
         \caption{$R2$}
         \label{fig:TauR2}
     \end{subfigure}
     \hfill
     \begin{subfigure}[b]{0.32\textwidth}
         \centering
         \includegraphics[width=0.8\textwidth]{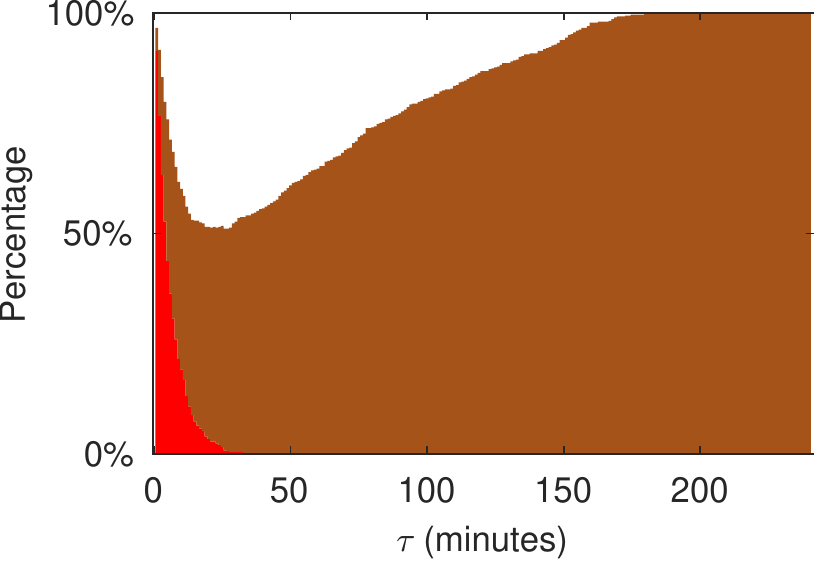}
         \caption{$R3$}
         \label{fig:TauR3}
     \end{subfigure}
        \caption{Percentage of undesired cases as a function of $\tau$ (out of 488 cases for R1, 715 for R2, and 501 for R3)}
        \label{fig:eva4}
\end{figure*}

\smallskip\noindent
In our evaluation, we assigned values $p1=0.95$, $p2=0.2$ and $p3=0.95$, and gave ranges and trends for the other parameters as shown in Table~\ref{parameterRange}. The range and value of model parameters were selected to ensure that the system-level properties would vary around the requirement, but can be based on prior knowledge in practice. Calculations were performed on a Macbook Pro with 2.3GHz Quad-core i7 CPU and 32 GB of RAM and all ``monitored'' parameters were updated every minute with a prediction horizon set for 240 minutes. 

\smallskip\noindent
\textbf{RQ1}: To answer this research question, we predicted the violation of a system-level requirement by propagating the predicted parameters 
within the prediction horizon via the PMC expressions presented in Table~\ref{Propsummary}. We first present one experiment in detail for illustration of the prediction results, and then we analyse the result from a large set of experiments. The detailed results are shown in Fig.~\ref{fig:eva1} where the x-axis shows the time in hours and the y-axis indicates the system-level property, for example, in Fig.~\ref{fig:RQ1R1}, the probability of completing the process with picking being successful. The bound from the requirement is shown as an orange dashed line. The ground truth for the system-level property is plotted as a solid blue line with predicted values as a dotted red line. The prediction horizon is highlighted in green with zero hours indicating the start. It can be seen that PRESTO is able to predict the violation of system-level requirements if there is any (R1 and R2), and predict the trend of the system level property if no violation is occurred within the prediction horizon (R3). The deterministic values of each monitored parameter at different time point are presented in Table~\ref{trendvalue} to show the tends. We evaluate the performance of PRESTO further by repeating the above experiment for 3000 times with randomly generated trends according to the specifications shown in Table~\ref{parameterRange}. For each experiment, we randomly chose two values that within the specified range for each parameter, and generated a set of values(600 value points for each parameters) that vary randomly between the two values. The unique trends of each parameter was created by sorting the value set in a descending or ascending order according to the specification in Table~\ref{parameterRange}. The first 360 values were used for parameter analysis and the last 240 values were used as the ground truth of prediction horizon. This evaluation can synthesise a large number of scenarios, including the violation of requirements before the prediction (1761 cases for R1, 1600 cases for R2, and 2209 cases for R3 out of 3000), violation of requirement within the prediction horizon (488 for R1, 715 for R2,and 501 for R3), and no violation of requirements during the period of analysis (751 for R1, 685 for R2, and 290 for R3), to minimise the impact of stochasticity in the experiments. The overall time span for a single run including prediction of time-series and evaluating the result is under one second, and the violation requirements can be observed uniformly across the prediction horizon. It is worth noting that during the 3000 runs of evaluating all three requirements, we did not observe any false positives (i.e., the prediction of a violation when one does not occur within the prediction horizon) or false negatives (i.e., a violation within the prediction horizon that was not predicted), which can be due to the combination effect of parameter trends, prediction horizon, and model structure. The violation of requirements before the prediction can be addressed by a reactive adaption method~\cite{calinescu2014adaptive} which is not in the scope of this work. We evaluated prediction errors, calculated as the difference in time between the predicted and actual violation of the requirement, for those experiments that have violation of requirement within the prediction horizon. The histograms together with fitted distributions are shown in Fig.~\ref{fig:eva2}. Our preliminary results show the prediction errors are normally distributed with mean values between -0.6 and 0.47 minutes (for R2 and R3 respectively), showing no real bias. The standard deviations of between 7.91 (for R3) and 11.4 (for R1) suggest that PRESTO is likely to predict system-level disruptions accurately.    


\begin{table}[]
\caption{Deterministic parameter values at three time points}
\label{trendvalue}
\sffamily
\vspace{-2mm}

\begin{small}
\begin{tabular}{p{1.2cm}p{0.5cm}p{0.5cm}p{0.5cm}p{0.5cm}p{0.5cm}p{0.5cm}p{0.5cm}p{0.5cm}}
\toprule
     {} & $\alpha$  & $\beta$ & $t_0$ &$t_1$& $t_2$ & $e_0$&$e_1$&$e_2$ \\ \midrule
now-360 & 0.98 & 0.01 & 1.04 & 10.01 &8.90 &3.30&2.40&0.30 \\ \midrule
now & 0.88&0.12&11.6&14.6&11.6&4.03&2.77&2.84 \\ \midrule
now+240 & 0.80&0.19&19.8&17.9&13.9&4.49&2.99&4.49 \\ \bottomrule
\end{tabular}
\end{small}

\vspace*{-2mm}
\end{table}

\begin{table}[]
\caption{Ranges and trends for the model parameters}
\label{parameterRange}
\sffamily

\vspace{-2.5mm}
\begin{small}
\begin{tabular}{p{0.8cm}p{2.2cm}p{1.2cm}p{1.2cm}p{1.2cm}}
\toprule
     {} & $\alpha$  & $\beta$ & $t_0,t_1,t_2$ & $e_0,e_1,e_2$ \\ \midrule
range & [0.7,0.99] & [0.01, 0.2] & [1s, 30s] & [0.3J, 4.5J] \\ \midrule
trend & constant or monotonically increasing & \multicolumn{3}{c}{constant or monotonically decreasing} \\ \bottomrule
\end{tabular}
\end{small}

\vspace*{-2mm}
\end{table}


\smallskip\noindent
\textbf{RQ2}: To evaluate the impact of noise on the prediction of disruptions, we added different levels of noise to the generated parameter trends using normal distributions with zero mean but different standard deviations as shown in Table~\ref{distribution}. The maximum value of $\alpha$ and $\beta$ are capped to one, and the minimum values for $t_1, t_2 and t_3$ and $e_1, e_2, and e_3$ are set to be 1s and 0.3J, respectively. The experiment was repeated 1000 times for each noise level to obtain their distributions as shown in Fig.~\ref{fig:eva3}. The distribution with zero mean (in blue), observed in all three cases, is obtained using trends with no added noise (level 0) and provides a baseline. As noise is added, the mean error shifts to the right due to noise spikes violating the requirement before the predicted value, which only considers violations caused by the underlying trend, here increasing or decreasing monotonically. The noise level also impacts the error variance, becoming larger as the noise level increases. These early stage results suggest PRESTO may be sensitive to noise, and that the integration of a de-noising module could be worthwhile.       


\begin{table}[]
\caption{Standard deviations of noise levels from \textbf{RQ3}}
\label{distribution}
\vspace{-2.5mm}

\sffamily
\begin{small}
\begin{tabular}{p{1.01cm}p{0.81cm}p{0.81cm}p{0.81cm}p{0.81cm}p{0.81cm}p{0.95cm}}
\toprule
     {} & level 0 & level 2  & level 4 & level 6 & level 8 & level 10 \\ \midrule
$\alpha$, $\beta$ & 0 & 0.02 & 0.04 & 0.06 & 0.08 & 0.1\\ \midrule
\textbf{$t_0,t_1,t_2$} & 0 & 2 & 4 & 6 & 8 & 10\\ \midrule
\textbf{$e_0,e_1,e_2$} & 0 & 0.6 & 1.2 & 1.8 & 2.4 & 3\\ \bottomrule
\end{tabular}
\end{small}

\vspace{-3mm}
\end{table}

\smallskip\noindent
\textbf{RQ3}: While our preliminary evaluation does not cover the end-to-end use of PRESTO within the MAPE-K loop of self-adaptive systems, we suggest how PRESTO predictions can be exploited by the planning step of the MAPE-K loop. It is clear that adaptation decisions need to be triggered at an appropriate time. In our example, the robot can trigger the decision any time between \textit{now} and predicted violation of requirements. We use a user specified variable, $\tau$, to determine the time to trigger the adaptation decision as max(\textit{now},$(t_{p} - \tau)$), where $t_{p}$ is the predicted time that the violation will occur. If  $\tau$ is  too large, the value $(t_{p} - \tau)$ could represent a time point in the past, in which case, the robot would need to make a reactive adaptation at \textit{now}, possibly generating additional adaptation cost. In addition, a large value of $\tau$ could lead to frequent unnecessary adaptations that affect the performance of the robot. On the other hand, if $\tau$ is too small, particularly when smaller than the prediction errors, the violation is more likely to happen before being predicted, resulting in high cost from recovery. To facilitate the selection of $\tau$, we illustrate the impact of this parameter in terms of the percentage of undesired cases out of 488 for R1, 715 for R2 and 501 for R3 that could occur using the same experimental data used for RQ1 (Fig.~\ref{fig:eva4}). Here $(t_{p}-\tau)\leq60 min$ is considered too large and $\tau<abs(t_{p}-t_{ref})$ is considered too small, where $t_{ref}$ is the time of the actual violation of the requirement. The three plots in Fig.~\ref{fig:eva4} show a consistent trend in that the number of undesirable cases is extremely high (up to 100\%) when $\tau$ is either close to zero or close to its maximum value. Our results show that $\tau$ has a sweet spot around 30-min as the percentage of undesired cases goes down as low as 50\%. However, if the violation of requirements is associated with a significant cost, a slightly larger $\tau$ is preferred.
\

\section{Related work}
\label{sec:relatedwork}
Self-adaptive systems can proactively or reactively adapt to changes~\cite{becker2012model}. While both adaptation techniques have shown good results in handling changes and uncertainties~\cite{arcelli2020exploiting}, proactive adaptation 
is particularly important when restoring compliance with requirements after they were violated incurs high costs~\cite{poladian2007leveraging} or latency~\cite{moreno2015proactive}.

Proactive adaptation relies on the prediction of current execution and environmental parameters to identify potential violations and to apply adaptations before these violations occur. While some research in this area concentrated on the improvement of prediction accuracy~\cite{metzger2018incremental,metzger2019proactive,camara2020quantitative}, others investigated possible limitations, such as the trade-off between early prediction and accuracy~\cite{metzger2020triggering}, the impact of adaptation latency~\cite{camara2014stochastic} and the impact of limited computational resources~\cite{quin2019efficient}. 

Related to PRESTO, the research from~\cite{moreno2015proactive} integrates probabilistic model checking within a latency-aware proactive adaptation method to determine the best adaptation strategy. The resulting method allows optimal adaptation decisions to be made over the prediction horizon with consideration of the stochastic behaviour of the environment. While this method can be applied at runtime, the entire model needs to be updated when the changes are observed. In contrast, our approach utilises parametric model checking, in which the same PMC expression can be re-used, with only its re-evaluation for new parameter values is required at runtime.  

\section{conclusion}
\label{sec:conclusion}

We have presented a new approach for the prediction of system-level disruptions (PRESTO) which is intended both for use in the analysis step and to support the planning step in the MAPE-K loop. PRESTO uses parametric model checking to establish functional relationships between system and/or environmental parameters and the system-level property. Time-series analysis is used to identify trends in the parameters from which values can be predicted and then propagated via the PMC expressions to assess their impact on system-level properties and to identify potential disruptions. Our preliminary evaluation suggests that PRESTO can accurately predict the system-level disruptions with no false positives or false negatives observed in our experiments. 


The benefit of a modular architecture means that the techniques used in PRESTO can be substituted with alternatives according to needs and new technologies can be added to extend functionality. We plan to integrate a de-noising module in future work as evaluation showed noise to have significant impact on the prediction results. Furthermore, more detailed evaluations need to be carried out with different pDTMCs and parameter trends to identify the limitations of PRESTO.

\section*{Acknowledgement}

This project funded by the UKRI project EP/V026747/1 `Trustworthy Autonomous Systems Node in Resilience'.

\clearpage
\bibliographystyle{plain}
\bibliography{reference}
\end{document}